\documentclass[final,5p,times,letterpaper,twocolumn]{elsarticle}

\usepackage[utf8]{inputenc}
\usepackage{hyperref}

\newcommand{\beq}{\begin{equation}}
\newcommand{\eeq}{\end{equation}}
\newcommand{\beqa}{\begin{eqnarray}}
\newcommand{\eeqa}{\end{eqnarray}}
\newcommand{\bfig}{\begin{figure}\begin{center}}
\newcommand{\efig}{\end{center}\end{figure}}
\newcommand{\btab}{\begin{table}\begin{center}}
\newcommand{\etab}{\end{center}\end{table}}

\journal{arXiv.org}

\begin{document}
\begin{frontmatter}

  \title{Advanced Techniques for Scientific Programming and
    Collaborative Development of Open Source Software Packages at the
    International Centre for Theoretical Physics (ICTP)}

  \author[ictp]{Ivan Girotto\corref{cor1}}
  \author[ictp,temple]{Axel Kohlmeyer\corref{cor1}}
  \author[ictp,duram]{David Grellscheid}
  \author[psc]{Shawn T. Brown}

  \address[ictp]{The Abdus Salam, International Centre for Theoretical
    Physics (ICTP) Strada Costiera 11, 34014 Trieste, Italy}
  \address[temple]{Institute for Computational Molecular Science,
    Temple University, Philadelphia, Pennsylvania 19122, United
    States}
  \address[duram]{Institute for Particle Physics Phenomenology, Durham University, South Rd, Durham DH1 3LE, UK}
  \address[psc]{Pittsburgh Supercomputing Center, Carnegie Mellon University, Pittsburgh, PA, USA} 
  \cortext[cor1]{Corresponding authors: \url{igirotto@ictp.it}, \url{akohlmey@ictp.it}}

  \begin{abstract}

    A large number of computational scientific research projects make
    use of open source software packages. However, the development
    process of such tools frequently differs from conventional
    software development; partly because of the nature of research,
    where the problems being addressed are not always fully
    understood; partly because the majority of the development is
    often carried out by scientists with limited experience and
    exposure to best practices of software engineering. Often the
    software development suffers from the pressure to publish
    scientific results and that credit for software development is
    limited in comparison.

    Fundamental components of software engineering like modular and
    reusable design, validation, documentation, and software
    integration as well as effective maintenance and user support tend
    to be disregarded due to lack of resources and qualified
    specialists. Thus innovative developments are often hindered by
    steep learning curves required to master development for legacy
    software packages full of \emph{ad hoc} solutions. The growing
    complexity of research, however, requires suitable and
    maintainable computational tools, resulting in a widening gap
    between the potential users (often growing in number) and
    contributors to the development of such a package.

    In this paper we share our experiences aiming to improve the
    situation by training young scientists, through
    disseminating our own experiences at contributing to open source
    software packages and practicing key components of software
    engineering adapted for scientists and scientific software
    development. Specifically we summarize the outcome of the
    ``Workshop in Advanced Techniques for Scientific Programming and
    Collaborative Development of Open Source Software Packages'' run
    at the Abdus Salam International Centre for Theoretical Physics in
    March 2013, and discuss our conclusions for future efforts.
  \end{abstract}

  \begin{keyword}
    Modular Software Design \sep Scientific Programming  \sep Collaborative Software Development \sep Open Source Scientific Software Packages
  \end{keyword}

\end{frontmatter}

\section{Introduction}
\label{sec:intro}

Since it was founded in 1964 by the late Nobel Laureate Abdus Salam,
the International Centre for Theoretical Physics (ICTP) has been a
driving force behind global efforts to advance scientific expertise in
the developing world. Under the governance of UNESCO and IAEA, ICTP
seeks to accomplish its mandate by providing scientists from
developing countries with the continuing education and skills that
they need to enjoy long and productive careers. One of the means to
achieve this goal is to organize workshops where young scientists from
both, the developing and the developed world learn side-by-side about
state of the art research and methodology.

The current strategic plan of the ICTP contains a particular
commitment for building and fostering resources and competences in
scientific software development and high-performance computing (HPC).
This applies to the infrastructure and staff at the ICTP as well as
its educational programs, and has resulted in the formation of a small team
specializing in development and execution of educational programs
for scientific and high-performance computing.

The complexity of current computational scientific research, especially
in scientific domains that require productivity through high-performance
computing, favors using flexible software packages or frameworks that
provide domain specific scripting, modular software design and code
reuse.  Often large parts of the work can be implemented in a script
language with time critical operations offloaded to compiled language
extensions or wrappers to external libraries. This way scientists can
realize new projects often without having to write and maintain a
complete application, but can focus on small domain specific modules and
at the same time leverage improvements to the framework that are
contributed by experts in other domains. In step with this, modern
multi-core architectures with vectorized floating point units as well as
accelerators like GPUs require a much higher level of programming
expertise to be effectively utilized for scientific computations.

For about the last 20 years, it was possible to implement efficient
scientific applications with just one programming language (often some
variant of Fortran) and using a rather minimal subset of the Message
Passing Interface (MPI) standard for parallelization. Many scientific
applications have matured into packages by adding new features and are
used by ever growing user communities. The growing demand for
computational resources was easily satisfied through more efficient
processors with higher clock rates and faster interconnects, without
significant changes to the overall structure of the software.

In light of the current disruptive changes in computer hardware
requiring a refactoring if not a complete rewrite of many codes, the
challenges to developers of scientific software are manyfold: they need
to understand the inner workings of new hardware, need to learn new
tools, libraries and programming paradigms to effectively exploit
this new hardware, need to embrace and support an ever
growing (non-technical) user community, need to handle increasingly
complex science and algorithms, and all of this frequently without
any formal training or funding for specifically these tasks.
Correspondingly, these challenges have to be addressed from multiple
directions. In the following we will focus on the aspects of developing
suitable training events and raising awareness in the research community
for supporting and educating qualified and motivated scientists to become
scientific software developers.

The ICTP ``Workshop in Advanced Techniques for Scientific Programming and
Collaborative Development of Open Source Software Packages'' was conceived
to address those two key issues through sharing experiences of
scientists involved in developing scientific
software packages and building and practicing fundamental skills in creating,
maintaining, and enhancing modular, reusable, and extensible software. Emphasis
is also given to practice working in a team with distributed responsibilities
and using modern collaborative software development and management tools
like distributed source code management, facilities for
validation and embedding documentation.

The workshop was planned as a three week activity with two weeks of general
training, followed by a third week dedicated to training of (future) 
developers of a particular software package. Leading expert developers
of that software are invited to show how the topics of the first two weeks
of the workshop can be translated to real world scientific software development
and also to get students 
to discuss with the expert developers, how they can improve the
adoption of the previously practiced methodologies.
For the inaugural workshop in 2013, the selected software package
was Quantum ESPRESSO (QE)~\cite{Quantum-Espresso}, since there already
exists a strong connection between QE developers, the ICTP and many 
potential workshop participants, which simplified planning and the content
development process. 

In the following sections we describe the event and our motivations and
observations about the program in detail. Section 2 will cover the generic
software development part of the first two weeks and features the
Quantum ESPRESSO Developer Training in the third week. The final part
of the document presents the authors' conclusions and raises questions
for further discussion.

\section{The Workshop Experience}
\label{sec:workshop}

The 2013 ``Workshop in Advanced Techniques for Scientific Programming and
Collaborative Development of Open Source Software Packages''
 was held March 11th to 22nd 2013 in Trieste at the ICTP with 
43 attendees from 25 countries (11 faculty and 32 participants)
with financial support of the ICTP for executing the workshop
and supporting scientists from developing countries.

The course was specifically targeted at computational scientists who are
already participating or want to participate in the development of
scientific software packages or plan to start such a project by
themselves. While many of such packages are grounded in HPC, the
workshop touched the topics of optimization and parallelization only
marginally and focused on software design, software
engineering practices, and collaborative development. Through lectures
and practical exercises participants were introduced to challenges of
complex scientific problems, modern computer hardware architectures,
modular software design, and collaborative software tools. This was
all done under consideration of the boundary conditions of scientific
research, where formal training in programming and software engineering
is inconsistent at best, and also skill sets and experience vary
significantly between different collaborators on a software project.
Since effective use and understanding of such methods and tools
requires practice, typical days were scheduled with about 3-4 hours of
lectures and 4-5 hours of practical demonstrations and exercises plus
additional opportunities for informal presentations and discussions.

The first week of the school focused on concepts and introduction of
the tools themselves, whereas the second week was dominated by
having groups of 4-6 participants work on small projects requiring the
use of tools and software design patterns introduced in the preceding
week. These projects comprised either refactoring of a small example
code in C or Fortran or developing a small software project from scratch.
Key topics of the workshop were: the design of modular and object
oriented applications, building applications with multiple levels of
abstraction that can use a scripting language like python at the top
level which is augmented in time critical parts with C/C++ or Fortran
subroutines, efficient use of tools for unit and regression testing,
debugging, profiling, documentation and source code management, and
building applications with support for accelerators and vectorization.
The topics and materials were directed at participants with
intermediate level knowledge in programming and physics.

\section{The Quantum ESPRESSO Developer Training}
\label{sec:qe-dev}

Around 37 participants from 18 different countries participated to the
Quantum ESPRESSO Developer Training (QE-Dev): 4 directors, 3
additional speakers, and 30 attendees. The QE-Dev was organized within
the same budget available for the first two weeks. 20 participants
have attended all three weeks of the activity, 10
of which were financially supported by ICTP. The first three days were
split into a morning session of direct lectures and an afternoon
practical session. The last day was fully dedicated to direct
lectures. Directors and speakers were actively participating in the
hands-on sessions as teaching assistants to guarantee adequate
support.

QE is an integrated suite of Open Source computer codes for electronic
structure calculations and materials modeling at the nanoscale. Like
the majority of community codes, it has a complex structure that has
been incrementally developed for more than two decades. Today the QE
distribution is composed of around half a million code lines, mostly
written in FORTRAN 90, and it includes a number of different
binaries. In spite of such complexity, it is considered a fundamental
tool for scientific research within the Condensed Matter community at
ICTP. A sizeable number of ICTP activities has been dedicated in the
past to the usage of QE.  The objective of the QE-Dev is to increase
the knowledge in the community of the different aspects of
development. For the first three days the morning lectures were aimed at creating the ground needed to face the hands-on
sessions. This included methodology and best practice of community
development: how to best contribute to both the improvement and the
maintenance of the package. Overall the first three days of the
workshop targeted fundamental aspects such as:
\begin{enumerate}
\item learning the most commons components (Modules, variables, routines) of the QE package, including an overview of the parallel approach;
\item learning how to perform standard tasks like wave-function I/O, calculating scalar product of 2 wave-functions, apply the Hamiltonian operator to a wave-function;
\item learning how to develop a Post-Processing Tool based on existing QE components.
\end{enumerate}
The last day was entirely dedicated to direct talks with particular focus on advanced topic such as: the LDA+U Implementation and Atomic Wavefunctions~\cite{LDA+U}, the Phonon package, the QE version for hybrid systems equipped with NVIDIA GPUs~\cite{QE-GPU} and a vision in depth of the levels of parallelism, including a detailed introduction on how to perform large-scale simulation~\cite{QE-Userguide,QE-EXX} .

\section{Conclusions}
\label{sec:conclusion}
Taking into consideration that the concept of the workshop was new
and significantly different from previous activities organized and
executed by the directors and staff, the directors are
unanimous in the appraisal of the event as extremely successful. We
attribute this to the following reasons:
\begin{enumerate}
\item Our excellent and dedicated workshop staff, all of whom
  performed admirably during a long and demanding event. Their efforts
  and time commitment far exceeded our requirements and expectations
  and were absolutely critical for making the event as successful as
  it was.
\item Efficient, professional, and prompt services provided by ICTP
  staff of the Adriatico Guest House along with the extensive help
  from the workshop secretary.
\item Careful selection of the participants. The selection was based
  primarily on answers to a provided self-evaluation questionnaire,
  which was rated on a demonstrated need for the course, scientific
  accomplishments (relative to their experience), recommendations and
  the appropriate skill level. This resulted in a very motivated group
  of participants, most of whom were very actively involved in both
  practical exercises and lectures. As a result, the morale of the
  workshop was very high despite the fact that the programming covered
  12 hours each day for both weeks and included Saturday morning.
\end{enumerate}
The co-scheduling with the QE-Dev resulted in a bias toward
participants with a background of condensed matter physics,
nevertheless other areas of computational physics were well
represented. The school proceeded without any disruptions and due to
the advanced level of experience, participants quickly managed to
adjust to the tasks of the practical sessions and actively
participated in lectures with competent and useful questions. We see
the largest accomplishment of the workshop in the successful execution
of the group projects. It quickly became visible, that scientists are
in general not prepared for collaborative software development and
thus their experience of their own difficulties and successes while
working on the group project is likely to significantly improve their
ability to collaborate; a skill that is becoming highly important in
computational sciences. The fact that all groups managed to present
tangible achievements at the end of the workshop and were highly
engaged in their respective projects demonstrated that we found a good balance
between the complexity of the project (medium) and the difficulty of
the physics.

While this workshop benefits from experiences made during previous
schools and workshops on high-performance computing, it contained a
significant amount of new teaching material and a different conceptual
focus and as a result, the program and topics will have to be refined
for future events of a similar kind. To that end, an evaluation of
the workshop was performed through an online form. Overall the
responses indicate a high approval rating and indicate some areas that
the directors also identified as needing improvement and are indicative
of the range of pre-existing skills. While requiring a minimum level
of experience to participate, the overall level of difficulty was
aimed a bit more toward the lower end of the distribution of skills
explaining some of the less favorable comments. On the other hand
directors could take advantage of the skill set distribution in the
group project phase by assigning participants to groups so that the
diversity of skills was represented in each group and added to the
challenge and learning experience.

The large number of applicants coming on their own funding and the
large number of applicants that were interested in co-scheduled QE-Dev
indicate that the workshop as such is addressing a need that is not
easily satisfied and that for future installments the co-scheduling
should be expanded into other large application packages with
relevance to the ICTP and affiliated communities.  A successful event
of this kind requires a large degree of commitment from the staff,
which must not only be knowledgeable, but also care about teaching and
handing their own skills to the developing world. The positive
feedback in this respect from participants as well as from tutors and
lecturers is an extremely gratifying experience. For the teaching
assistants who were at or close to the graduate student level, the
workshop provided an opportunity to hone their skills in managing a
small group of scientists, an experience that cannot be easily had
otherwise until much later in their career. Any future version of this
workshop should continue in this tradition.

The QE-Dev was a new initiative not only for ICTP but also for the
world-wide scientific community behind this package. Despite several
years of experience by both directors and speakers, gathered during
the activity of code development, a significant effort was requested
to teach how a given problem is coded and structured within the
package. Only a small amount of teaching material was available for such
a purpose, especially for the hands-on sections where an entire program
had to be developed from scratch.

The directors of the QE-Dev consider the outcome of the training school as
the starting point on which it is possible to base and build 
similar Developer Training world-wide, in analogy to the already established
users' schools of QE. An on-line
evaluation form was proposed for evaluation of the QE-Dev,
too. The answers received about the
relevance of most of the topic presented were extremely positive. On the other hand, lower
enthusiasm has been expressed in regards to how those topic were
presented.  This experience suggests that the following actions be
taken for the preparation of a similar tutorial in the future: more
carefully prepared lectures; longer hands-on sessions; distribution of
a written text with the supporting documentation; the selection of a
more homogeneous group of participants.  In conclusion, we report of a
successful event that definitely allows scientist to meet together for
a common and challenging purpose: the development of a scientific
community code.

We are extremely pleased with the way the overall `Workshop in
Advanced Techniques for Scientific Programming and Collaborative
Development of Open Source Software Packages'' has worked out. The
very high dedication of the participants, the majority of whom were
very attentive and motivated to learn, was an extremely rewarding
experience.

The first experience fed enthusiasm to directors and stakeholders for
pursuing such activities in the future while changing the target of
the scientific community for what concerns the last week. For 2014,
LAMMPS~\cite{LAMMPS} is the selected software package. We aim to
promote initiatives that can further motivate the participation of
scientists that see the potential for opportunities in new software
developments. The presented model of the workshop will be extended,
and more open to fresh developers. To complete the event a symposium
will be organized. The main topic for the symposium is intended to be
the presentation of research works that are based on either usage or
new development for the LAMMPS software package or similar. The
extension to less restricted sessions should aid in reducing the
distance between scientists and what we consider an important component
in the future of scientific research: a collaborative software
development.  We aim to further export the model introduced in ICTP
by this workshop, while extending it to a co-organized event outside of
ICTP, with particular interest to developing countries.

We aim to build the ground to attract possible contributors from
developing countries to the development of open source scientific
packages. The relatively small cost for investments needed to create
such conditions can increase possibilities for scientists from
developing countries to leverage their research work within the
world-wide scientific community. In step with the ICTP strategic
mission, scientists from developed countries can be attracted to
participate in such unique events aiding the possibility
of contact with local scientific communities. 

Following up on the previous development training organized in early
2013 at ICTP, scientists from the QE community have organized an
advanced training for the December 2013. The event is aimed at a
selected group of developers that will undertake an (on-line)
pre-course to guarantee an adequate preparation to both theoretical
and practical aspects of the workshop. The main part of the laboratory
session will see participants working closely with world-class experts
with the final goal of implementing additional new features into the
official distribution of Quantum ESPRESSO. 

\section*{Acknowledgments}
We thank the ICTP for providing an excellent venue and the opportunity
to conduct this type of event.  We thank ICTP and the Quantum
ESPRESSO Foundation for providing the ground to make such event
possible.  Especially, we would like to thank Prof. Stefano De
Gironcoli (SISSA/DEMOCRITOS), Prof. Paolo Giannozzi (University of
Udine), Dr. Paolo Umari (University of Padova), Dr. Carlo Cavazzoni
(CINECA) as well as Prof. Andrea dal Corso (SISSA) for their
availability and Dr. Emine Kucukbenli (EPFL) for the essential
contribution and scientific support for the QE-Dev. Ms. Rosa del Rio,
secretary of the workshop, made organizing the workshop
straightforward and gave the directors the much needed freedom to
focus on the program.




\begin{thebibliography}{00}
\bibitem{Quantum-Espresso}
  P. Giannozzi, S. Baroni, N. Bonini, M. Calandra, R. Car, C. Cavazzoni,
  D. Ceresoli, G. L. Chiarotti, M. Cococcioni, I. Dabo, A. Dal Corso,
  S. de Gironcoli, S. Fabris, G. Fratesi, R. Gebauer, U. Gertsmann,
  C. Gougoussis, A. Kokalj, M. Lazzeri, L. Martin-Samos, N. Marzari,
  F. Mauri, R. Mazzarello, S. Paolini, A. Pasquarello, L. Paulatto,
  C. Sbraccia, S. Scandolo, G. Sclauzero, A. P. Seitsonen, A. Smogunov,
  P. Umari and R. M. Wentzcovitch,
  J. Phys.: Condens. Matter \textbf{21}, 395502 (2009);
  \url{http://www.quantum-espresso.org}

\bibitem{LDA+U}
M. Cococcioni and Stefano de Gironcoli, "Linear response approach to the calculation of the effective interaction parameters in the LDA+U method", Physical Review B 71, 035105 (2005)

\bibitem{QE-GPU} 
Spiga, F. \& Girotto, I. phiGEMM: A CPU-GPU Library for Porting Quantum ESPRESSO on Hybrid Systems, 2012 20th Euromicro International Conference on Parallel, Distributed and Network-Based Processing (PDP), doi: 10.1109/PDP.2012.72 Publication Year: 2012 , Page(s): 368 - 375. IEEE Conference Publications

\bibitem{QE-Userguide}Quantum ESPRESSO user guide. [Online]. Available:http://www.quantum-espresso.org/users-manual/

\bibitem{QE-EXX} N. Varini, D. Ceresoli, L. Martin-Samos, I. Girotto, C. Cavazzoni, "Enhancement of DFT-calculations at petascale: Nuclear Magnetic Resonance, Hybrid Density Functional Theory and Car-Parrinello calculations", 2013, Comp. Phys. Comm, doi: 10.1016/j.cpc.2013.03.003, Volume 184, Issue 8, August 2013, Pages 1827-1833

\bibitem{LAMMPS}
S. Plimpton, Fast Parallel Algorithms for Short-Range Molecular Dynamics, J Comp Phys, 117, 1-19 (1995)



\end{thebibliography}
\end{document}